\documentclass[11pt]{iopart}

\usepackage{iopams}
\usepackage{amsfonts}
\usepackage{amssymb}

\usepackage{graphicx}
\graphicspath{{pics/}} 

\usepackage[justification=justified,singlelinecheck=true]{caption}
\usepackage[caption=false]{subfig}
\usepackage{hyperref}

\usepackage{booktabs}
\usepackage{calc}
\usepackage{array}
\newcolumntype{L}[1]{>{\raggedright\let\newline\\\arraybackslash\hspace{0pt}}m{#1}}
\newcolumntype{C}[1]{>{\centering\let\newline\\\arraybackslash\hspace{0pt}}m{#1}}
\newcolumntype{R}[1]{>{\raggedleft\let\newline\\\arraybackslash\hspace{0pt}}m{#1}}

\usepackage{multirow}

\usepackage{algpseudocode}
\usepackage{algorithm}

\usepackage{needspace}

\usepackage{iopmath_my}
\usepackage{iopext_my}

\newlength{\colwidthA}
\newlength{\colwidthB}
\newlength{\colwidthC}
\newlength{\colwidthD}
\newlength{\colwidthE}
\newlength{\lastcolwidth} 
\newlength{\floatwidth}

\newcommand{\stabGroup}[1][G]{\mathcal{S}_{#1}}

\begin{document}
    %
\title{Determining X-chains in graph states}
\author{
Jun-Yi Wu
,
Hermann Kampermann
,
Dagmar Bru\ss
}
\address{
Institut f\"ur Theoretische Physik III, Heinrich-Heine-Universit\"at D\"usseldorf, D-40225 D\"usseldorf, Germany
}

%


        %
\begin{abstract}
The representation of graph states in the X-basis as well as the calculation of graph state overlaps can efficiently be performed by using the concept of X-Chains [Phys. Rev. A 92(1) 012322].
We present a necessary and sufficient criterion for X-chains and show that they can efficiently be determined by Bareiss algorithm.
An analytical approach for searching X-chain groups of a graph state is proposed.
Furthermore we generalize the concept of X-chains to so-called Euler chains, whose induced subgraphs are Eulerian.
This approach helps to determine if a given vertex set is an X-chain and we show how Euler chains can be used in the construction of multipartite Bell inequalities for graph states.
%
\end{abstract}
        %

\tableofcontents
\maketitle
    %

    %
\section{Introduction}
\label{sec::introduction}
Graph states are states of multipartite quantum system which are created by two-qubit control-Z gates.
Their entanglement is rich in structure, such that they can be employed in measurement based quantum computation \cite{RaussendorfBriegel2001-05,RaussendorfBriegel2002-10,RaussendorfBrowneBriegel2003-08,RaussendorfPHDThesis2003}.
Graph states can be represented in the stabilizer formalism, which indicates application in quantum error correction \cite{SchlingemannWerner2001-QECGraph, BellHTMWRarity2014-ExpGSQEC}.
Besides, they can also be exploited in quantum secret sharing \cite{ChenLo2007-QCryptGHZ, MarkhamSanders2008-GSforQSS, BellHMWRTame2014-ExpGSQSS}.
\par
The number of entangling gates needed in the preparation of a graph state is in general equal to the edge number of its underlying graph.
In practical application, noise in the entangling gates leads to imperfect graph states.
For this situation, an error model for the preparation of graph states via repeat until success control-Z gates \cite{LimBeigeKwek2005-RUS,LimBBKKwek2006-RUS,BeigeLimKwek2007-RUS} was studied in \cite{WuRKSKMBruss2014-05},
where the so-called randomized graph states were introduced to represent the end products of such a process.
\par
To witness entanglement of randomized graph states, one needs to calculate overlaps of two graph states.
These overlaps are determined by \emph{X-chain} groups (Def. \ref{def::X-chains}), which are the vertex subsets corresponding to the stabilizer of a graph state consisting of only $\sigma_X$ Pauli operators \cite{WuKampermannBruss2015-XChainFactor}.
X-chains also characterize graph states in the X-basis \cite{WuKampermannBruss2015-XChainFactor}.
%
%
\bigskip
\par
In this paper, we study the search of X-chains of a given graph state.
This search can be efficiently implemented by the Bareiss algorithm \cite{Bareiss1968-AlgoForGaussElimi}.
We derive an X-chain criterion (Theorem \ref{theorem::criterion_for_X-chains}) and introduce the so-called \emph{X-chain fragments} (Def. \ref{def::X-chain_fragment}).
X-chain fragments are vertex subsets, which are the basic constituents of X-chains.
Via the X-chain criterion, and X-chain fragment merging and removal (Proposition \ref{prop::criterion_for_redundant_fragments}, \ref{prop::merging_of_intersecting_fragments}, \ref{prop::criterion_for_fragment_merging_1} and \ref{prop::criterion_for_fragment_merging_2}), one obtains the X-chain groups of certain types of graph states, e.g. star graph states, linear cluster states, cycle graph states and complete graph states (Corollary \ref{coro::X-chains_and_bias_degree_of_S_n}, \ref{coro::X-chains_and_bias_degree_of_L_n}, \ref{coro::X-chains_and_bias_degree_of_C_n} and \ref{coro::X-chains_and_bias_degree_of_K_n}).
\par
As a generalization of X-chains, Euler chains are introduced as the vertex subsets, whose induced subgraphs are Euler graphs (Def. \ref{def::Eulertian_chain}).
The graph state stabilizers induced by a certain type of Euler chains
were employed in graph state quantum repeaters \cite{EppingKampermannBruss2015}.
Euler chains are easier to verify than X-chains.
We show the relation between Euler chains and X-chains (Corollary \ref{coro::X-chains_and_Euler_chains} and \ref{coro::X-chains_of_Eulerian_gs}), so that one can verify if a vertex subset is an X-chain through the criterion for Euler chains.
Furthermore, Euler chains can be applied in the construction of multipartite Bell inequalities for graph states according to the approach in Ref. \cite{TothGuhne2005-EntDetectStabilizer}(Theorem \ref{theorem::MP_Bell_ineq_n_Eulerian_chains}).
\bigskip
\par
This paper is organized as follows: we first review the definitions and properties of graph states and X-chains in section \ref{sec::graph_states_and_X-chains}.
Then we study the search of X-chains in section \ref{sec::search_of_X-chains}, where an X-chain criterion is proposed.
An analytical approach based on X-chain fragments is derived in section \ref{sec::X-chain_fragments}, and the X-chain groups of special graph states are given in section \ref{sec::X-chain_groups_of_special_graphs}.
Furthermore, we introduce Euler chains and show their relation to X-chains in section \ref{sec::Euler_chains_n_X-chains}.
At the end, we propose the application of Euler chains in the construction of multipartite Bell inequalities in section \ref{sec::multi_Bell_ineq_Euler_chains}.
\section{Review: graph states and X-chains}
\label{sec::graph_states_and_X-chains}
Graph states can be defined in the stabilizer formalism.
Each vertex $i$ is associated with a \emph{stabilizer generator},
\begin{equation}
g_{i}=\sigma_{x}^{(i)}\sigma_{z}^{(N_{i})}.
\end{equation}%
Here, $N_{i}$ is the neighborhood of the vertex $i$, $\sigma_{x}^{(i)}$ stands for the Pauli operator $\sigma_{x}$ applied to the $i$-th qubit, and $\sigma_{z}^{(N_{i})}:=\bigotimes_{j\in N_{i}}\sigma_{z}^{(j)}$ is the tensor product of $\sigma_{z}$ applied to the neighbors of the $i$-th qubit.
An $n$-vertex graph state $|G\rangle$ is an $n$-qubit state stabilized by $g_{i}$, i.e.,
\begin{equation}
g_{i}|G\rangle=|G\rangle,\text{for all } i=1,...,n.
\end{equation}
The $n$ graph state stabilizer generators, $g_{i}$, generate the whole stabilizer group $\stabGroup$ of $|G\rangle$ via the product operation.
The group $\stabGroup$ is Abelian and contains $2^{n}$ elements. These $2^{n}$ stabilizers uniquely represent a graph state on $n$ vertices.
Each graph state stabilizer can be induced from stabilizer generators by its corresponding vertex subset $\xi$.
We introduced therefore in Ref. \cite{WuKampermannBruss2015-XChainFactor} the so-called induced stabilizers as follows.
\begin{definition}[\cite{WuKampermannBruss2015-XChainFactor} Induced stabilizers]
\label{def::induced_stabilizer}
\Needspace*{8\baselineskip}
Let $G$ be a graph on vertices $V_{G}=\left\{1,2,...,n\right\}  $. Let $\xi=\left\{  i_{1},\cdots,i_{m}\right\}  $ be a subset of $V_{G}$.
We call the product of all $\left\{g_{i}\right\}  _{i\in\xi}$, i.e.
\begin{equation}
s_{G}^{(\xi)}:=\prod_{i\in\xi}g_{i},\label{eq::induced_stabilizer}%
\end{equation}
the \emph{$\xi$-induced stabilizer} of $|G\rangle$.
Here, $g_{i}$ is the stabilizer of $|G\rangle$ associated with vertex $i$.
\end{definition}
The $\xi$-induced stabilizer has the explicit form
\begin{equation}
s_{G}^{(\xi)}=\pi_{G}(\xi)  \sigma_{x}^{(\xi)}\sigma_{z}^{(c_{\xi})},
\label{eq::induced_stabilizer_math_formula}%
\end{equation}
where $c_{\xi}$ is called the \emph{correlation index} of $\xi$ in $|G\rangle$ and
defined as the \emph{$2$-modulo neighbourhood} of $\xi$, i.e.%
\begin{equation}
c_{\xi}:=N_{i_{1}}\Delta N_{i_{2}}\Delta\cdots\Delta N_{i_{m}},%
\label{eq::correlation_index_math_formula}%
\end{equation}
where $\Delta$ is the symmetric difference operation of sets\footnote{The symmetric difference of two sets is defined as $\xi_1\Delta\xi_2:=(\xi_1\cup\xi_2)\setminus(\xi_1\cap\xi_2)$.}, and $N_{i_j}$ is the neighbourhood of the vertex $i_j$ in the whole graph $G$.
Correspondingly, $\xi$ is called the X-resource of the correlation index $c_{\xi}$.
In Eq. \eqref{eq::induced_stabilizer_math_formula}, $\pi_{G}\left(  \xi\right)  $ is the \emph{stabilizer-parity} of $\xi$, which is defined as%
\begin{equation}
\pi_{G}\left(  \xi\right)  :=(-1)^{\left\vert E(G[\xi])\right\vert
},\label{eq::G-parity_formula}%
\end{equation}
where $\left\vert E(G[\xi])\right\vert $ is the edge number of the $\xi
$-induced subgraph $G[\xi]$,
i.e. the stabilizer-parity of $\xi$ is the parity of the edge number $|E(G[\xi])|$.
\bigskip
\par
There are special graph state stabilizers $s_G^{(\xi)}$ consisting of solely Pauli-X operators, i.e.
\begin{equation}
s_{G}^{(\xi)}=\pi_{G}\left(  \xi\right)  \sigma_{x}^{(\xi)}.
\end{equation}
From these stabilizers, the representation of graph states in the X-basis \cite{WuKampermannBruss2015-XChainFactor} and overlaps of two graph states can be determined.
The vertex subsets, which induce such special graph state stabilizers, are called X-chains.
\begin{definition}[\cite{WuKampermannBruss2015-XChainFactor} X-chains]
\label{def::X-chains}
  Let $G$ be a graph on vertices $V$. A set of vertices $\xi\subseteq V$ is an X-chain, if its correlation index is empty, i.e. $c_{\xi}=\emptyset$.
\end{definition}
The set of X-chains is shown to be a group with the group operation $\Delta$ (symmetric difference of sets)\cite{WuKampermannBruss2015-XChainFactor}.
An X-chain group is denoted by $\braket{\Gamma_G}$ with its generating set being written as $\Gamma_G$.
Its elements $\gamma\in\Gamma_G$ are called X-chain generators.
The quotient group $\braket{\mathcal{K}_G}:=\mathcal{P}_G/\braket{\Gamma_G}$ is called the correlation group of the graph state $\ket{G}$.
\par
The \emph{(Z-)bias degree} of a graph state $\ket{G}$, $\beta(\ket{G}):=\braket{+|G}$, is the overlap of $\ket{G}$ and the plus state $\ket{+}^{\otimes n}$,
where $\ket{+}=(\ket{0}+\ket{1})/\sqrt{2}$.
The value $2^n\beta(\ket{G})$ is the difference of the numbers of positive and negative amplitudes of the Z-basis states in the superposition of $\ket{G}$.
The bias degree is equal to \cite{WuKampermannBruss2015-XChainFactor}
\begin{equation}
\beta(|G\rangle)=\frac{1}{2^{\left(  n-\left\vert \Gamma_{G}\right\vert
\right)  /2}}\prod_{\gamma\in\Gamma_{G}}\delta_{\pi_{G}(\gamma)}^{1},
\label{eq::bias_degrees_of_graph_states}
\end{equation}
where $\delta$ is the Kronecker-delta.
The overlap of two graph states $\ket{G}$ and $\ket{H}$ is equal to
\begin{equation}\label{eq::graph_state_overlaps}
  \braket{G|H} = \beta(\ket{G\Delta H}),
\end{equation}
where $G\Delta H$ is the graph symmetric difference\footnote{The graph symmetric difference of the graphs $G$ and $H$ is defined as $G\Delta H=(V, E_G\Delta E_H)$ with $V=V_G=V_H$.}.
For a given graph state, if there is an X-chain with negative parity, then the bias degree of the graph state is zero.
Such a graph state has the same number of positive and negative amplitudes in the superposition of the Z-basis states, and hence is called \emph{Z-balanced} graph state.
\section{The search of X-chains}
\label{sec::search_of_X-chains}
The binary vector\footnote{The binary vector of a vertex subset $\xi$ is given by $(i_1,...,i_n)$ with $i_j=1$ if $j\in\xi$, otherwise $i_j=0$} of correlation index $i^{(c_{\xi})}$ can be mapped from the binary vector of vertex subset $i^{(\xi)}$ via the adjacency matrix $A_G$ of $G$, i.e.
\begin{equation}
  i^{(c_{\xi})}=A_G i^{(\xi)} \pmod 2.
\end{equation}
According to the definition of X-chains (Def. \ref{def::X-chains}), $i^{(c_{\gamma})}=0$, the X-chain generators $\gamma$ are then the basis of the modulus-$2$ kernel space of $A_G$.
The kernel of $A_G$ can be calculated via Gaussian Elimination.
It can be implemented via the Bareiss algorithm \cite{Bareiss1968-AlgoForGaussElimi}, which is polynomial with respect to the vertex number $n$.
This allows to efficiently determine the X-chains of general graph states.
\bigskip
\par
For the analysis of the X-chain groups of certain graph states, we derive the following theorem.
\begin{theorem}[Criterion of X-chains]
\label{theorem::criterion_for_X-chains}
\Needspace*{12\baselineskip}
    A set of vertices $\xi$ is an X-chain of a graph state $\ket{G}$, if and only if the number of edges between $\xi$ and any vertex $v\in V_G$ is even, which means%
    \begin{equation}
    \left\vert E_G(\xi,v)\right\vert = 0 \pmod{2},
    \label{eq::criterion_for_X-chains_1}
    \end{equation}
    for all $v \in V_G$, and is equivalent to
    \begin{equation}
    \left\vert N_{v}\cap\xi\right\vert = 0 \pmod{2}.
    \label{eq::criterion_for_X-chains_2}
    \end{equation}
\begin{proof}
Let $\xi$ be an X-chain, that means its correlation index $c_{\xi}=\emptyset$.
According to Eq. \eqref{eq::correlation_index_math_formula}, it must hold%
\begin{equation}
\Delta_{j\in\xi}N_{j}(G)=\emptyset.
\label{eq::proof_theorem_X-chain_cri}%
\end{equation}
This equation holds, if and only if for any vertex $v\in V_{G}$, $v$ is contained in an even number of neighborhoods $N_{j}$ within $j\in\xi$.
That means $\left\vert N_{v}\cap\xi\right\vert $ is even
for all $v\in V_{G}$.
\end{proof}
\end{theorem}%
%
\begin{table}[th] 

\newlength{\Xgraphwidth}
\setlength{\Xgraphwidth}{0.16\columnwidth}
\newlength{\thirdcolwidth}
\setlength{\thirdcolwidth}{0.15\columnwidth}
\newlength{\fourthcolwidth}
\setlength{\fourthcolwidth}{0.15\columnwidth}
\setlength{\lastcolwidth}{0.9\columnwidth-3\Xgraphwidth-\thirdcolwidth-\fourthcolwidth}

\def\gwidth{0.7}
\begin{tabular}
[c]{
|C{\Xgraphwidth}
|C{2\Xgraphwidth}
|C{\thirdcolwidth}
|C{\fourthcolwidth}
|C{\lastcolwidth}
|
}%
\toprule

Graph $G$ & X-chains $\braket{\Gamma_G}$ & X-chain generators $\Gamma_G$ & Stabilizer parity $\pi_{G}(\xi)$ & Bias degree $\langle+^{\otimes n}|G\rangle $\\
\midrule
%
\includegraphics[width=\linewidth]{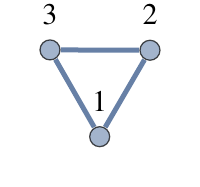}
\newline
$K_3$
&%
\makebox[0.9\linewidth]{
\begin{tabular}{C{0.5\linewidth}C{0.5\linewidth}}%
  $\xi_{1}=\emptyset$ & $\xi_{2}=\{1,2,3\}$\\
  \includegraphics[width=\gwidth\linewidth]{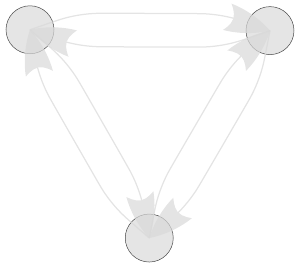} & \includegraphics[width=\gwidth\linewidth]{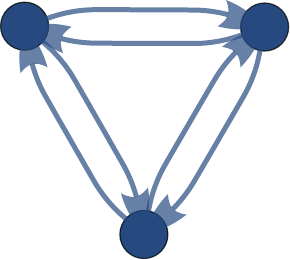} \\
\end{tabular}
}
& \{\{1,2,3\}\} & $\pi_{G}(\xi_{1})=1$ \newline $\pi_{G}(\xi_{2})=-1$ & $0$\\

\midrule
\includegraphics[width=\linewidth]{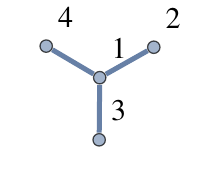}
\newline
$S_4$
&%
\makebox[0.9\linewidth]{
\begin{tabular}{C{0.5\linewidth}C{0.5\linewidth}}%
  $\xi_{1}=\emptyset$ & $\xi_{2}=\{2,3\}$\\
  \includegraphics[width=\gwidth\linewidth]{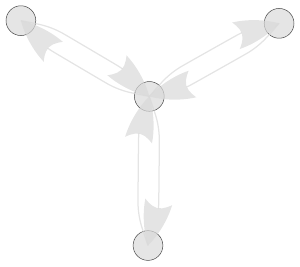} & \includegraphics[width=\gwidth\linewidth]{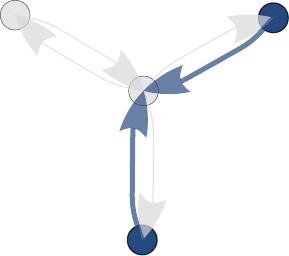} \\
  $\xi_{3}=\{2,4\}$ & $\xi_{4}=\{3,4\}$\\
  \includegraphics[width=\gwidth\linewidth]{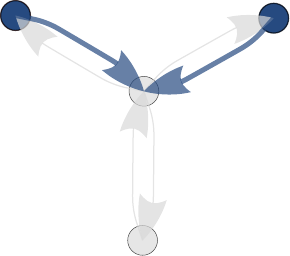} & \includegraphics[width=\gwidth\linewidth]{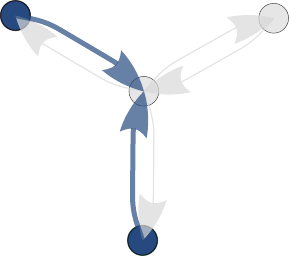} \\
\end{tabular}
}
& \{\{2,3\},\{2,4\}\} & $\pi_{G}(\xi_{i})=1$, for all $\xi_{i}$ & $1/2$ \\
\bottomrule
\end{tabular} 
\caption{\colorfig X-chain groups of simple graphs: the graphs shown under the column ``X-chains'' demonstrate the criterion in Theorem \ref{theorem::criterion_for_X-chains}.
Vertices of X-chains are highlighted.
We draw out-going arrows from the vertices of an X-chain $\xi$ pointing to their neighbors.
One observes that the number of incident arrows on each vertex is even, i.e. $|N_v\cap \xi|=0 \pmod 2$ for all $v\in V_G$.
}
\label{table::x-chain_group_of_graph_state_eg}
\end{table}
Table \ref{table::x-chain_group_of_graph_state_eg} shows two examples of graph states with their X-chain groups, X-chain generators and bias degrees.
With this theorem, the X-chain groups of star graph states $\ket{S_n}$, whose underlying graph has a center vertex with $n-1$ neighbors and all the others have the center vertex as their only neighbor (see Table \ref{table::X-chains_of_special_graphs_Sn}), can be directly determined as follows.
\begin{corollary}[X-chain generators of star graphs]
\label{coro::X-chains_and_bias_degree_of_S_n}
\Needspace{4\baselineskip}
The X-chain generating set of a star graph state $|S_{n}\rangle$ on the vertices $V_{S_n}=\left\{1,...,n\right\} $ with vertex $1$ as its center (see Table \ref{table::X-chains_of_special_graphs_Sn}) is $\Gamma_{S_n}=\left\{\{i,i+1\}\right\}_{i=2,...,n-1}$.
The bias degree of $|S_n\rangle$ is $\beta(\ket{S_n})=1/2$.
\begin{proof}
  For all $i=2,...,n-1$, $N_j\cap \{i,i+1\}=2\delta_{1j}$, which means $\{i,i+1\}$ are X-chains, i.e. $\left\{\{i,i+1\}\right\}_{i=2,...,n-1}\subseteq\braket{\Gamma_{S_n}}$.
  \par
  On the other hand, if a vertex subset $\xi$ is an X-chain, as a result of the X-chain criterion (Theorem \ref{theorem::criterion_for_X-chains}), it must satisfy $1\notin\xi$ and $|\xi\cap N_1|$ being even, which means $\xi\subseteq N_1$ and $|\xi|=0\pmod 2$.
  Since all vertex subsets $\xi\subseteq N_1$ with even cardinality can be generated by $\left\{\{i,i+1\}\right\}_{i=2,...,n-1}$ via the symmetric difference $\Delta$ of sets,
  it holds $\braket{\left\{\{i,i+1\}\right\}_{i=2,...,n-1}}\supseteq\braket{\Gamma_{S_n}}$.
  \par
  As a result $\Gamma_{S_n}=\left\{\{i,i+1\}\right\}_{i=2,...,n-1}$.
\end{proof}
\end{corollary}%
According to this corollary the X-chain generators of the $5$-vertex star graph state $\ket{S_5}$ in Table \ref{table::X-chains_of_special_graphs_Sn} is $\{\{2,3\},\{3,4\},\{4,5\}\}$.
\begin{table}[ht!]
  \centering
  \begin{tabular}
[c]{|C{0.04\textwidth}|C{0.2\textwidth}|C{0.1\textwidth}|C{0.4\textwidth}|C{0.1\textwidth}|}
\toprule
$|G\rangle$ & & $n$ & $\Gamma_{G}= \left\{  \{i,i+1\}  :i=2,...,n-1\right\}  $ & $\beta(\ket{G})$\\
\midrule
$|S_{n}\rangle$ & \includegraphics[width=0.9\linewidth]{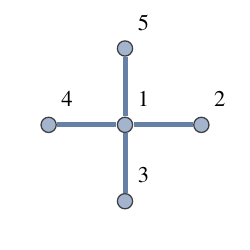} & $5$ & $ \{\{2,3\},\{3,4\},\{4,5\}\}$ & $\frac{1}{2}$\\
\bottomrule
\end{tabular} 
  \caption{X-chains of star graph states.}
  \label{table::X-chains_of_special_graphs_Sn}
\end{table}
\par
The search of X-chain groups via Theorem \ref{theorem::criterion_for_X-chains} is a combinatorial problem, in which the less number of combining fragments it has, the easier it can be solved.
Instead of a single vertex, we consider the fundamental constituents of X-chains, called \emph{X-chain fragments}, which will be defined in section \ref{sec::X-chain_fragments}.
Two criteria for merging of two X-chain fragments will be given.
Consequently, one can obtain the X-chain groups of certain special graph states, which will be studied in section \ref{sec::X-chain_groups_of_special_graphs}.
\subsection{X-chain fragment merging and removal}
\label{sec::X-chain_fragments}
\begin{figure}[th] 
  \centering
  \newlength{\gwidth}
\setlength{\gwidth}{0.45\columnwidth}
\begin{tabular}{
|C{0.45\columnwidth}
|C{0.45\columnwidth}
|
}
\toprule
\multicolumn{2}{|c|}{\includegraphics[width=1.5\gwidth]{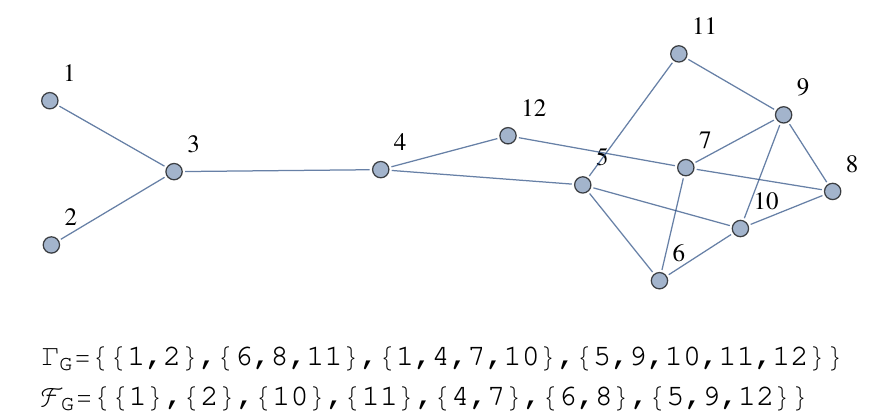}} \\
%
\midrule
\includegraphics[width=\gwidth]{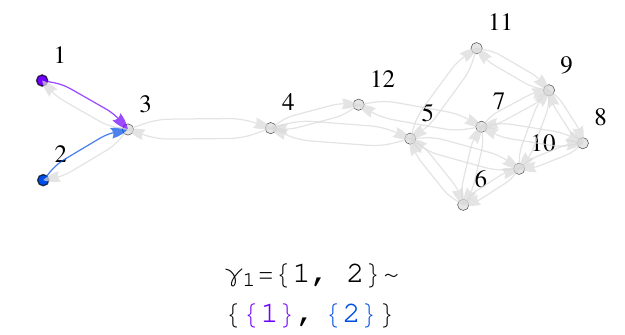} & \includegraphics[width=\gwidth]{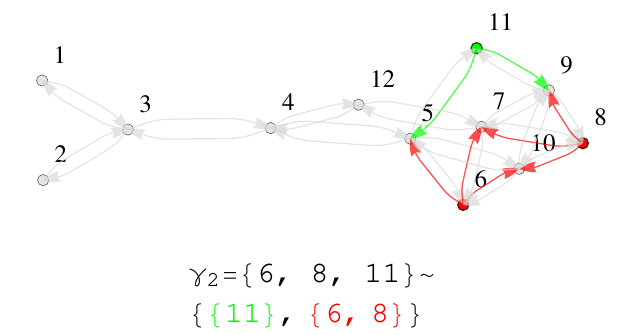} \\
\midrule
\includegraphics[width=\gwidth]{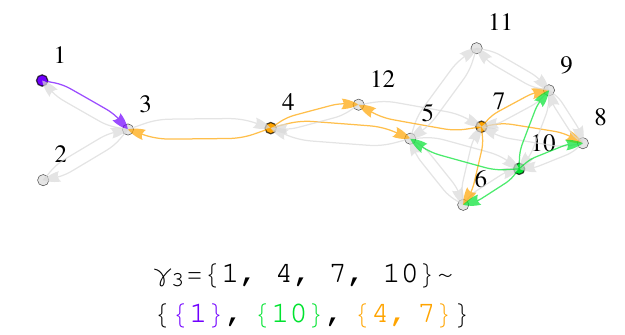} & \includegraphics[width=\gwidth]{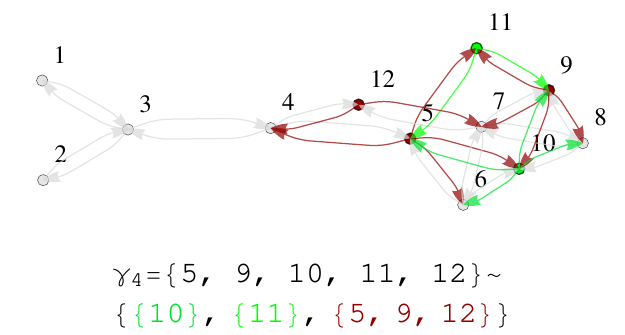} \\
\bottomrule
\end{tabular} 
  \caption{\colorfig An example of X-chain fragments: the fish-like graph has $4$ X-chain generators.
  The mutual intersection and exclusion of these generators form the set of X-chain fragments $\mathcal{F}_G$.
  The X-chain generators are certain combinations of X-chain fragments. The arrows show the fulfillment of the X-chain criterion (Theorem \ref{theorem::criterion_for_X-chains}) by these combinations.
 }
  \label{fig::X-chain_fragment_set_eg}
\end{figure}%
Consider the graph in Fig. \ref{fig::X-chain_fragment_set_eg}, one observes that the vertices $4$ and $7$ contribute to X-chains together as an integrity, which also holds for example for the vertex sets $\{6,8\}$ and $\{5,9,12\}$.
We call such bonds of vertices \emph{X-chain fragments}, which are defined as follows.%
\begin{definition}[X-chain fragments]
\label{def::X-chain_fragment}
\Needspace*{10\baselineskip}
A set of vertices $f=\left\{v_{1},...,v_{k}\right\}$ of a graph $G$ is an X-chain fragment, if $f$ is  contained by at least one X-chain, and it is either a subset of or has no common elements with all X-chains $\xi \in \braket{\Gamma_G}$ of the graph state $\ket{G}$, i.e.%
\[
f\cap\xi\in\left\{  \emptyset,f\right\} .
\]

\end{definition}%
As a result of this definition, one can exclude certain vertices from elements of X-chain fragments via the following proposition.
\begin{proposition}[Negative criterion for X-chain fragments]
\label{prop::criterion_for_redundant_fragments}
A vertex set $\alpha$ is not an X-chain fragment, if there is a vertex $v$, such that $N_v\subseteq \alpha$ and $|N_v|=1\pmod 2$.
\begin{proof}
Assume that $\xi$ is an X-chain containing $\alpha$, and there exists a vertex $v$ such that  $N_v\subseteq \alpha$ and $|N_v|=1\pmod 2$, then $N_v\subseteq \xi$ and the cardinality of $N_v\cap \xi = N_v$ is odd, which contradicts $N_v\cap \xi = 0\pmod 2$(Theorem \ref{theorem::criterion_for_X-chains} ).
As a result $f$ is not contained by any X-chain, hence is not an X-chain fragment.
\end{proof}

\end{proposition}%
For the example of the fish-like graph state in Fig. \ref{fig::X-chain_fragment_set_eg}, the vertex set $\{3\}$ is not an X-chain fragment, since $N_1\subseteq \{3\}$ and $|N_1|=1$.
A further result of the definition of X-chain fragments is that X-chain fragments with nonempty intersection can be merged into larger fragments iteratively until they are mutually disjoint.
\begin{proposition}[Merging of intersecting X-chain fragments]
\label{prop::merging_of_intersecting_fragments}
\Needspace*{6\baselineskip}
Let $f_{1}$ and $f_{2}$ be two X-chain fragments and $f_{1}\cap f_{2}\not=\emptyset $, then $f_{1}\cup f_{2}$ is an X-chain fragment.
\begin{proof}
See \ref{apdx::proofs_for_X-chains}.
\end{proof}
\end{proposition}
Non-intersecting X-chain fragments can also be merged according to the following two criteria.
\begin{proposition}[Criterion for X-chain fragment merging I]
\label{prop::criterion_for_fragment_merging_1}
\Needspace*{6\baselineskip}
Let $f_{1}$ and $f_{2}$ be two disjoint X-chain fragments.
If there exists a vertex $v$, such that $N_v\subseteq f_1\cup f_2$ and the cardinalities $|N_v\cap f_1|$ and $|N_v\cap f_2|$ are both odd,
then $f_1\cup f_2$ is an X-chain fragment.
\begin{proof}
See \ref{apdx::proofs_for_X-chains}.%
\end{proof}
\end{proposition}%
Consider the $5$-vertex linear cluster state in Table \ref{table::X-chains_of_special_graphs_Ln}, $\{1,3\}$ is an X-chain fragment according to this proposition, since $N_2\subseteq\{1\}\cup\{3\}$ and $|N_2\cap\{1\}|=|N_2\cap\{3\}|=1$.
Analogously, $\{3,5\}$ is an X-chain fragment.
As a result of Proposition \ref{prop::merging_of_intersecting_fragments}, $\{1,3,5\}$ is the remaining X-chain fragment after merging ($\{2\}$ and $\{4\}$ are not X-chain fragments according to Proposition \ref{prop::criterion_for_redundant_fragments}).
\begin{proposition}[Criterion for X-chain fragment merging II]
\label{prop::criterion_for_fragment_merging_2}
\Needspace*{14\baselineskip}
Let $f_{1}$ and $f_{2}$ be two disjoint X-chain fragments.
If there exist two vertices $v_{1}$ and $v_{2}$, such that%
\begin{equation}
N_{v_{1}}\setminus N_{v_{2}}=f_{1} \text{ and } N_{v_{2}}\setminus N_{v_{1}}=f_{2},
\end{equation}%
and the edge number between $v_i$ and $f_i$ is odd for $i=1,2$, i.e.
\begin{equation}\label{eq::prop_criterion_for_fragment_merging_cond}
|N_{v_i}\cap f_i|=1 \pmod 2,
\end{equation}
then $f_{1}\cup f_{2}$ is an X-chain fragment.
\begin{proof}
See \ref{apdx::proofs_for_X-chains}.
\end{proof}
\end{proposition}%
For the $5$-vertex complete graph state in Table \ref{table::X-chains_of_special_graphs_Kn} the vertex set $\{1,2\}$ is an X-chain fragment, since $N_1\setminus N_2 = \{2\}$, $N_2\setminus N_1 = \{1\}$, and $|N_1\cap\{2\}|=|N_2\cap\{1\}|=1$.
The same holds for any pair of vertices $\{i,j\}$ in any complete graph state with odd vertex number.
(For complete graph states with even vertex number there exist no X-chains, hence no X-chain fragments, see Corollary \ref{coro::X-chains_and_bias_degree_of_K_n}.)
\par
For certain graph states, after iterative fragment merging (Proposition \ref{prop::merging_of_intersecting_fragments}, \ref{prop::criterion_for_fragment_merging_1}, \ref{prop::criterion_for_fragment_merging_2}) and removal (Proposition \ref{prop::criterion_for_redundant_fragments}), it is possible to obtain a small number of X-chain fragments, such that the combinatorial problem in the X-chain criterion (Theorem \ref{theorem::criterion_for_X-chains}) becomes easy to solve.
Furthermore, if an X-chain fragment $f$ has an even number of edges between all vertex, then $f$ is an X-chain itself.
In the next section, we determine the X-chain groups of certain graph states via this approach.
\subsection{X-chain groups of special graph states}
\label{sec::X-chain_groups_of_special_graphs}
Linear cluster states $\ket{L_n}$, complete graph states $\ket{K_n}$ and cycle graph states $\ket{C_n}$ are special graph states, whose X-chain groups can be determined directly after fragment merging and removal (see section \ref{sec::X-chain_fragments}).
Their X-chain groups and Z-bias degrees are given in the following corollaries.
\begin{table}[ht!]
  \centering
  \begin{tabular}
[c]{|C{0.04\textwidth}|C{0.2\textwidth}|C{0.1\textwidth}|C{0.25\textwidth}|C{0.25\textwidth}|}
\toprule
$|G\rangle$ &  & $n$ & $\Gamma_{G}$ & $\beta(\ket{G})$ \\
\midrule
\multirow{2}{*}{$|L_{n}\rangle$} & \includegraphics[width=0.9\linewidth]{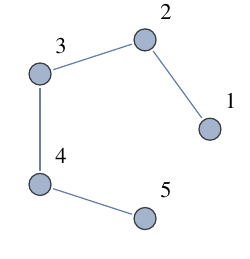} &
$2k-1$ & $\left\{  \left\{  1,3,...,2k+1\right\}  \right\}  $ & $2^{-(n-1)/2}$\\
\cline{2-5}
& \includegraphics[width=0.9\linewidth]{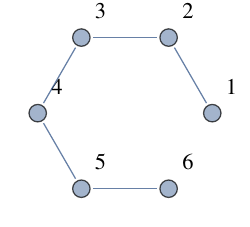} & $2k$ & $\emptyset$ & $2^{-n/2}$ \\
\bottomrule
\end{tabular} 
  \caption{X-chains of linear cluster graph states.}
  \label{table::X-chains_of_special_graphs_Ln}
\end{table}
\par
Linear cluster states $\ket{L_n}$ are graph states, whose vertices lie on a line with two end points and only the nearest vertices are connected by edges (see Table \ref{table::X-chains_of_special_graphs_Ln}).
Their X-chain groups are the following.
\begin{corollary}[X-chains and Z-bias degree of linear cluster states]
\label{coro::X-chains_and_bias_degree_of_L_n} \Needspace{12\baselineskip}
The X-chain generators and Z-bias degree of linear cluster states $|L_{n}\rangle$ with vertex set $V_{L_{n}}=\left\{ 1,...,n\right\}  $ (see Table \ref{table::X-chains_of_special_graphs_Ln}) depend on the parity of vertex number $n$.
\begin{enumerate}
\item If $n$ is odd, $\Gamma_{L_{n}}=\left\{  \gamma_{\mathrm{odd}}\right\}  $ with $\gamma_{\mathrm{odd}}=\left\{  1,3,...,n-2,n\right\}  $, and $\beta(|L_{n}\rangle)=2^{-\left(  n-1\right)  /2}$.

\item If $n$ is even, $\Gamma_{L_{n}}=\emptyset$, and $\beta(|L_{n}%
\rangle)=2^{-n/2}$.
\end{enumerate}
\begin{proof}
See \ref{apdx::proofs_for_X-chains}.
\end{proof}
\end{corollary}
\begin{table}[ht!]
  \centering
  \begin{tabular}
[c]{|C{0.04\textwidth}|C{0.2\textwidth}|C{0.15\textwidth}|C{0.2\textwidth}|C{0.25\textwidth}|}
\toprule
$|G\rangle$ &  & $n$ & $\Gamma_{G}$ & $\beta(\ket{G})$ \\
\midrule
\multirow{3}{*}{$|K_{n}\rangle$} & \includegraphics[width=0.9\linewidth]{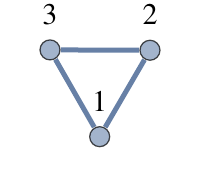} &
$4k+3,k\geq0$ & $\left\{  \left\{  1,...,n\right\}  \right\}  $ & $0$ \\
\cline{2-5}
& \includegraphics[width=0.9\linewidth]{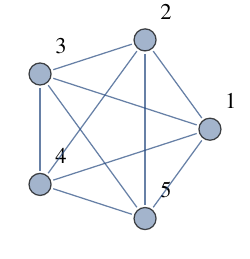} & $4k+1, k\geq1$ & $\left\{
\left\{  1,...,n\right\}  \right\}  $ & $2^{-(n-1)/2}$ \\
\cline{2-5}
& \includegraphics[width=0.9\linewidth]{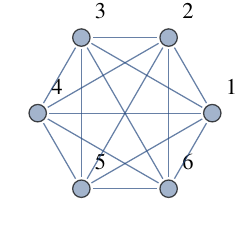} & $2k$ & $\emptyset$ &
$2^{-n/2}$ \\
\bottomrule
\end{tabular} 
  \caption{X-chains of complete graph states.}
  \label{table::X-chains_of_special_graphs_Kn}
\end{table}
\par
Complete graph states $\ket{K_n}$ are graph states, whose underlying graphs contain all possible edges between the vertices (see Table \ref{table::X-chains_of_special_graphs_Kn}).
Their X-chain groups are the following.
\begin{corollary}[X-chains and Z-bias degree of complete graph states]
\label{coro::X-chains_and_bias_degree_of_K_n}
\Needspace*{12\baselineskip}
The X-chain generators and Z-bias degree of complete graph states $|K_{n}\rangle $ with vertex set $V_{K_{n}}=\left\{ 1,...,n\right\} $ (see Table. \ref{table::X-chains_of_special_graphs_Kn}) depend on the parity of vertex number $n$.
\begin{enumerate}
\item If $n$ is odd, $\Gamma _{K_{n}}=\left\{ V_{K_{n}}\right\} $, and $\beta (|K_{n}\rangle )=0$ for $n=3 \pmod 4$, while $\beta (|K_{n}\rangle )=2^{-\left( n-1\right) /2}$ for $n=1 \pmod 4$.
\item If $n$ is even, $\Gamma _{K_{n}}=\emptyset $, and $\beta(|K_{n}\rangle )=2^{-n/2}$.
\end{enumerate}
\begin{proof}
See \ref{apdx::proofs_for_X-chains}.
\end{proof}
\end{corollary}
\begin{table}[ht!]
  \centering
  \begin{tabular}
[c]{|C{0.04\textwidth}|C{0.2\textwidth}|C{0.1\textwidth}|C{0.35\textwidth}|C{0.15\textwidth}|}
\toprule
$|G\rangle$ &  & $n$ & $\Gamma_{G}$ & $\beta(\ket{G})$ \\
\midrule
\multirow{2}{*}{$|C_{n}\rangle$} & \includegraphics[width=0.9\linewidth]{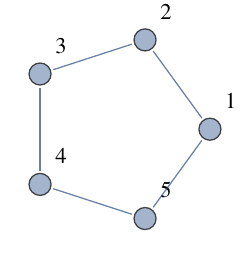} & $2k-1$ & $\left\{  \left\{  1,...,n\right\}  \right\}  $ & $0$ \\
\cline{2-5}
   & \includegraphics[width=0.9\linewidth]{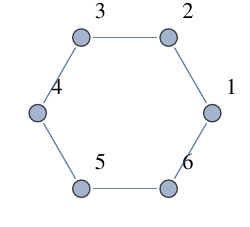} & $2k$ & $\left\{
\left\{  1,3,...,n-1\right\}, \left\{  2,4,...,n\right\}  \right\}  $ & $2^{-(n-2)/2}$ \\
\bottomrule
\end{tabular} 
  \caption{X-chains of cycle graph states.}
  \label{table::X-chains_of_special_graphs_Cn}
\end{table}
\par
Cycle graph states $\ket{C_n}$ are graph states, whose underlying graphs are connected closed paths (i.e. every vertex has degree $2$, see Table \ref{table::X-chains_of_special_graphs_Cn}).
Their X-chain groups are the following.
\begin{corollary}[X-chains and Z-bias degree of cycle graph states]
\label{coro::X-chains_and_bias_degree_of_C_n}
\Needspace{12\baselineskip}
The X-chain generators and Z-bias degree of cycle graph states $|C_{n}\rangle $ with vertex set $V_{C_{n}}=\left\{1,...,n\right\} $ (see Table. \ref{table::X-chains_of_special_graphs_Cn}) depend on
the parity of vertex number $n$.
\begin{enumerate}
\item If $n$ is odd, $\Gamma _{C_{n}}=\left\{ V_{C_{n}}\right\} $, and $%
\beta (|C_{n}\rangle )=0$.
\item If $n$ is even, $\Gamma _{C_{n}}=\left\{ \gamma _{\mathrm{odd}},\gamma_{\mathrm{even}}\right\} $, with $\gamma_{\mathrm{odd}}=\left\{ 1,3,...,n-1\right\} $ and $\gamma _{\mathrm{even}}=\left\{ 2,4,...,n\right\} $.
    The Z-bias degree is $\beta(|C_{n}\rangle )=2^{-(n-2)/2}$.
\end{enumerate}
\begin{proof}
See \ref{apdx::proofs_for_X-chains}.
\end{proof}
\end{corollary}
\section{Euler chains}
\label{sec::Euler_chains}
\subsection{Necessary condition for X-chain}
\label{sec::Euler_chains_n_X-chains}
\begin{figure}[th]
  \centering
  \subfloat[]{
   \def\cellwidth{0.27\textwidth}
\begin{tabular}{|C{\cellwidth}|C{\cellwidth}|C{\cellwidth}|}
  \toprule
  \includegraphics[width=0.9\linewidth]{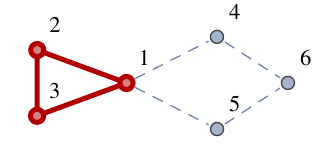} &
  \includegraphics[width=0.9\linewidth]{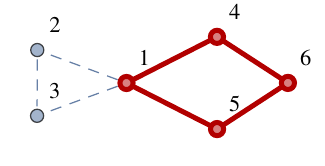} &
  \includegraphics[width=0.9\linewidth]{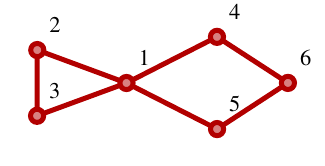} \\
  \bottomrule
\end{tabular}

   \label{fig::Euler_chain_eg_fish}
  }
  \\
  \subfloat[]{
   \def\cellwidth{0.17\textwidth}
\begin{tabular}{|C{\cellwidth}|C{\cellwidth}|C{\cellwidth}|C{\cellwidth}|C{\cellwidth}|}
  \toprule
  \includegraphics[width=\linewidth]{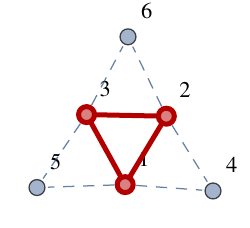} &
  \includegraphics[width=\linewidth]{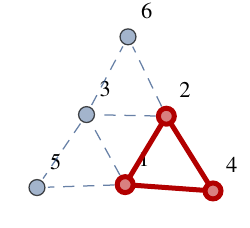} &
  \includegraphics[width=\linewidth]{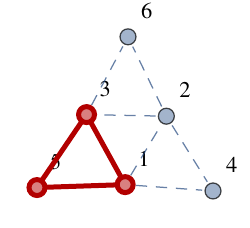} &
  \includegraphics[width=\linewidth]{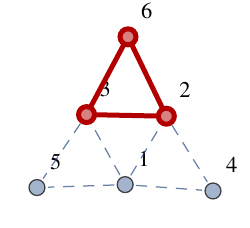} &
  \includegraphics[width=\linewidth]{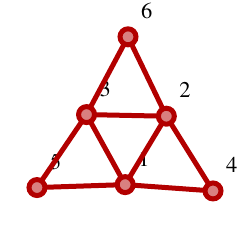} \\
  \bottomrule
\end{tabular}

   \label{fig::Euler_chain_eg_trig}
  }
  \caption{\colorfig Euler chains: the subgraph $G[\xi]$ induced by Euler chains $\xi$ are highlighted in bold and red.
  The correlation index $c_{\xi}$ and X-resource $\xi$ do not have common elements.
  (a) The Euler stabilizers corresponding to the Euler chains $\{1,2,3\}$ and $\{1,4,5,6\}$ are $\sigma_X^{\{1,2,3\}}\sigma_Z^{\{4,5\}}$ and $\sigma_X^{\{1,4,5,6\}}\sigma_Z^{\{2,3\}}$, respectively.
  The Euler chain $\xi=V_G$ is an X-chain according to Corollary \ref{coro::X-chains_of_Eulerian_gs}.
  (b) The Euler stabilizer corresponding to the Euler chains $\{1,2,4\}$, $\{1,3,5\}$ and $\{2,3,6\}$ are $\sigma_X^{\{1,2,4\}}\sigma_Z^{\{5,6\}}$, $\sigma_X^{\{1,3,5\}}\sigma_Z^{\{4,6\}}$ and  $\sigma_X^{\{2,3,6\}}\sigma_Z^{\{4,5\}}$, respectively.
  The Euler chain $\{1,2,3\}$ is an X-chain, while $V_G$ is also an X-chain according to Corollary \ref{coro::X-chains_of_Eulerian_gs}.
  }
  \label{fig::Euler_chain_eg}
\end{figure}

In this section, we introduce the so-called \emph{Euler chains}, which are more general and easier to verify than X-chains.
\begin{definition}[Euler chains and Euler stabilizers]
\label{def::Eulertian_chain}
  A vertex subset $\xi\subseteq V_G$ of a graph state $\ket{G}$ is an \emph{Euler chain}, if and only if its induced subgraph $G[\xi]$ is an Euler graph, i.e. the vertex degree $d_v(G[\xi])$ in $G[\xi]$ is even for all $v\in \xi$\footnote{Some authors refer by Euler graphs to connected graphs, which contain Euler cycles.}.
  If $G[\xi]$ is an empty graph, then $\xi$ is a \emph{trivial Euler chain}.
  The stabilizer induced by an Euler chain is an \emph{Euler stabilizer}.
\end{definition}
According to this definition, an Euler chain $\xi$ is trivial, if and only if the vertices inside $\xi$ are not neighbors to each other.
A single vertex is a trivial Euler chain, and hence the graph state stabilizer $g_i$ is a trivial Euler stabilizer.
The examples of the non-trivial Euler chains of the fish-like graph state and triangle graph state are highlighted in Fig. \ref{fig::Euler_chain_eg_fish} and \ref{fig::Euler_chain_eg_trig}, respectively.
Euler stabilizers are special graph state stabilizers consisting of no $\sigma_Y$ Pauli operators.
\begin{theorem}[Euler stabilizers of graph states]
\label{theorem::Eulertian_stabilizer}
\Needspace{8\baselineskip}
  Let $\xi$ be an Euler chain.
  The correlation index $c_{\xi}$, see Eq. \eqref{eq::correlation_index_math_formula}, does not intersect with $\xi$.
  That means the Euler stabilizer
  \begin{equation}\label{eq::theorem_Euler_stabilizer}
    s_G^{(\xi)} = \pi_G(\xi)\sigma_X^{(\xi)}\sigma_Z^{(c_{\xi})}
  \end{equation}
  does not contain $\sigma_Y$-Pauli operators.
\begin{proof}
  $\xi$ is an Euler chain, if and only if the vertex degree $d_v(G[\xi])$ is even for all $v\in \xi$, which means $N_v\cap \xi$ is even for all $v\in \xi$.
  For all $v'\in\xi$, $v'$ appears an even number of times in the sequence of the symmetric difference operation in Eq. \eqref{eq::correlation_index_math_formula}, which means $v'\notin c_{\xi}$.
  As a result, $\xi\cap c_{\xi}=\emptyset$.
  According to Eq. \eqref{eq::induced_stabilizer_math_formula}, $s_G^{(\xi)}$ does not contain $\sigma_Y$ operators.
\end{proof}
\end{theorem}
As a result of this theorem, an X-chain is certainly an Euler chain.
Hence being an Euler chain is a necessary condition for X-chains.
\begin{corollary}[X-chains and Euler chains]\label{coro::X-chains_and_Euler_chains}
  If the $\xi$-induced subgraph $G[\xi]$ is not an Euler graph, then $\xi$ is not an X-chain of the graph state $\ket{G}$.
\end{corollary}
To determine if a vertex subset is an Euler chain, one just needs to know the vertex degree $d_v(G[\xi])$.
The computational complexity of checking the vertex degrees of all vertices in $\xi$ is $O(|\xi|)$. This is much lower than the complexity of Bareiss algorithm exploited in the combinatorial condition of the X-chain criterion (Theorem \ref{theorem::criterion_for_X-chains}), which is $O(2 (\log n) n^5)$ with $n\ge|\xi|$ being the vertex number of the whole graph. Therefore, for a given vertex subset $\xi$ of a graph state $\ket{G}$ with large vertex number $n\gg|\xi|$, determining Euler chains is easier than X-chains.
%
%
\bigskip
\par
According to Theorem \ref{theorem::Eulertian_stabilizer}, if the underlying graph of a graph state is an Euler graph, then it has at least one X-chain, namely the whole vertex set $V_G$.
\begin{corollary}[X-chains of Euler graph states]
\label{coro::X-chains_of_Eulerian_gs}
  The vertex set $V_G$ is an X-chain of the graph state $\ket{G}$, if and only if $G$ is an Euler graph (i.e. all vertex degrees $d_v$ are even.)
\end{corollary}
For example, the graphs in Fig. \ref{fig::Euler_chain_eg_fish} and \ref{fig::Euler_chain_eg_trig} are both Euler graphs, their vertex set $\{1,2,3,4,5,6\}$ is therefore an X-chain of their corresponding graph states.
\subsection{Multipartite Bell inequalities}
\label{sec::multi_Bell_ineq_Euler_chains}
Graph state stabilizers, which contain no $\sigma_Y$ Pauli operators, were employed in graph state quantum repeaters in Ref. \cite{EppingKampermannBruss2015}.
There the graph state stabilizers in use are induced by trivial Euler chains, whose vertices are not neighbors to each other (see Def. \ref{def::Eulertian_chain}).
%
%
\par
On the other hand, non-trivial Euler chains can be applied in multipartite Bell inequalities.
In Ref. \cite{TothGuhne2005-EntDetectStabilizer}, entanglement witnesses constructed in the stabilizer formalism were proposed.
By evaluating the bound of the witness operators in LHV models, one can derive Bell inequalities.
As a special stabilizer state, the projectors of graph states can be employed as a Bell operator, i.e.
\begin{equation}\label{eq::graph_states_as_Bell_operators}
  B_G:=\projector{G} = \sum_{\xi\subseteq V_G} s_G^{(\xi)}.
\end{equation}
Since $\bra{G}B_G\ket{G}=1$, if the upper bound $C_{\mathrm{LHV}}(B_G)$ of the expectation value $\braket{B_G}_{\mathrm{LHV}}$ in LHV models is smaller than 1, then the quantum bound $\bra{G}B_G\ket{G}$ violates the Bell inequality $\braket{B_G}_{\mathrm{LHV}} \leq C_{\mathrm{LHV}}(B_G)$.
In general, the stabilizers of graph states in the righthand side of Eq. \eqref{eq::graph_states_as_Bell_operators} contain $\sigma_X$, $\sigma_Y$ and $\sigma_Z$ Pauli-operators.
The corresponding Bell measurement setting contains then $n$-partite local measurements with $3$ inputs (measurement directions) and $2$ outcomes.
Via selecting Euler stabilizers from the whole stabilizer group, one can construct Bell inequalities associated with local measurements with only $2$ measurement directions, i.e. $\sigma_X$ and $\sigma_Z$.
Note that not every Bell operator constructed by Euler stabilizers has an LHV bound that can be violated by a quantum state.
We propose an approach of deriving Bell inequalities for graph states from Euler chains in the following theorem.
\begin{theorem}[Bell inequalities for graph states and Euler chains]
\label{theorem::MP_Bell_ineq_n_Eulerian_chains}
Let $|G\rangle$ be a graph state.
If $\xi$ is an Euler chain of $\ket{G}$ with negative stabilizer parity, the following multipartite Bell operator%
\begin{equation}
B_{\xi}=\sum_{i\in\xi}g_{i}+s_{G}^{(\xi)}
\end{equation}
with $g_{i}=\sigma_{x}^{(i)}\sigma_{z}^{(N_{i})}$ is dichotomic.
Its expectation value in a local hidden variable model is bounded by
\begin{equation}
  -|\xi|+(-1)^{|\xi|+1} \leq \braket{B_{\xi}}_{LHV} \leq |\xi|-1,
\end{equation}
while its quantum bound $\left\vert \xi\right\vert +1$ is reached by the graph state $\ket{G}$.
\begin{proof}
  See \ref{apdx::proofs_for_X-chains}.
\end{proof}
\end{theorem}
For an Euler graph state one can obtain a simple Bell inequality as follows.
\begin{corollary}[Bell inequalities for Euler graph states]
\label{coro::Bell_ineq_for_Euler_gs}
  An Euler graph state $\ket{G}$ with odd edge number violates the Bell inequality
  \begin{equation}\label{eq::Bell ineq_for_Euler_gs}
    \Braket{B_{V_G}}_{\mathrm{LHV}}=\Braket{\sum_{i\in V_G} g_i - \sigma_X^{(V_G)}}_{\mathrm{LHV}}\leq |V_G|-1
  \end{equation}
  with the quantum value $|V_G|+1$.
\end{corollary}
This corollary follows directly from Theorem \ref{theorem::MP_Bell_ineq_n_Eulerian_chains}.
The odd edge number is required to fulfill the prerequisite of negative stabilizer parity in Theorem \ref{theorem::MP_Bell_ineq_n_Eulerian_chains}.
Since every cycle graph with $2k+1$ vertices and the complete graph with $4k-1$ vertices is an Euler graph and has odd edge number ($k\in \mathbb{N}$),
Eq. \eqref{eq::Bell ineq_for_Euler_gs} is a Bell inequality for the graph states $\ket{C_{2k+1}}$ and $\ket{K_{4k-1}}$.
For instance, with $G=K_3$ being the $3$-vertex complete graph (see Table \ref{table::X-chains_of_special_graphs_Kn}) one obtains
\begin{equation}\label{eq::Mermin_ineq_C3_ineq}
  \Braket{B_{V_G}}_{\mathrm{LHV}}=\Braket{
  \sigma_X^{\{1\}}\sigma_Z^{\{2,3\}}
  +\sigma_X^{\{2\}}\sigma_Z^{\{1,3\}}
  +\sigma_X^{\{3\}}\sigma_Z^{\{1,2\}}
  -\sigma_X^{\{1,2,3\}}}_{\mathrm{LHV}} \leq 2,
\end{equation}
which is identical to the Mermin inequality for $3$-qubit systems \cite{Mermin1990-BellIneq}.
The graphs in Fig. \ref{fig::Euler_chain_eg_fish} and \ref{fig::Euler_chain_eg_trig} are both Euler graphs with odd edge number.
Hence, for example, the Bell inequality for the graph state in Fig. \ref{fig::Euler_chain_eg_fish} can be derived with the whole vertex set $V_G=\{1,...,6\}$, i.e.
\begin{equation}\label{eq::Bell_ineq_Fish_ineq}
  \Braket{B_{V_G}}_{\mathrm{LHV}} \leq 5
\end{equation}
with
\begin{align}\label{eq::Bell_operator_Fish}
  B_{V_G} = &
  \sigma_X^{\{1\}}\sigma_Z^{\{2,3,4,5\}}
  +\sigma_X^{\{2\}}\sigma_Z^{\{1,3\}}
  +\sigma_X^{\{3\}}\sigma_Z^{\{1,2\}}
  +\sigma_X^{\{4\}}\sigma_Z^{\{1,6\}}
  +\sigma_X^{\{5\}}\sigma_Z^{\{1,6\}}
  +\sigma_X^{\{6\}}\sigma_Z^{\{4,5\}}
  \\ &
  -\sigma_X^{\{1,2,3,4,5,6\}}.
\end{align}

\section{Conclusion and outlook}
%
%
%
In this paper, we studied the method of searching X-chains of graph states.
In general, the X-chain group of a given graph state $\ket{G}$ can be calculated via the $2$-modulus kernel of the adjacency matrix $A_G$, which is computable with the Bareiss algorithm in polynomial time.
\par
For the analysis of certain types of graph states, a criterion for X-chains was derived in Theorem \ref{theorem::criterion_for_X-chains}, which states that the edge number between an X-chain and each vertex must be even.
As a result of this theorem, the X-chain groups and Z-bias degree of star graph states are given analytically.
\par
X-chain fragments are vertex subsets, which are the elemental constitutions of X-chains (Definition \ref{def::X-chain_fragment}).
By merging X-chain fragments iteratively together with exclusion of redundant vertex sets,
one obtains a small number of X-chain fragments for certain graph states, such that the search of X-chain generators via the criterion for X-chains (Theorem \ref{theorem::criterion_for_X-chains}) can be facilitated.
For linear cluster states $\ket{L_n}$, cycle graph states $\ket{C_n}$ and complete graph states $\ket{K_n}$ their X-chain generators are easily determined by fragment merging and removal.
\bigskip
\par
Furthermore, the concept of Euler chains was introduced as a generalization of X-chains (Def. \ref{def::Eulertian_chain}).
Their induced graph stabilizers do not contain $\sigma_Y$ Pauli operators (Theorem \ref{theorem::Eulertian_stabilizer}).
Given a vertex subset $\xi$ in a graph state, it can not be an X-chain, if it is not an Euler chain.
To determine whether $\xi$ is an Euler chain is easier than whether it is an X-chain,
which is especially useful for vertex subsets in graph states with large number of vertices.
We showed that an Euler graph state has at least one X-chain, namely its whole vertex set $V_G$.
We showed that Euler chains can be used for the construction of multipartite Bell inequalities for graph states (Theorem \ref{theorem::MP_Bell_ineq_n_Eulerian_chains}).
\bigskip
\par
Due to the complex structure of 2D cluster states, an analytic expression for their X-chains is still an open question.
Also the reverse problem, that given a vertex subset $\xi$, what are the graph states $\ket{G}$ that possess $\xi$ as an X-chain, is open.
Note that Corollary \ref{coro::X-chains_and_Euler_chains} might lead to a solution of this problem.
Apart from Theorem \ref{theorem::MP_Bell_ineq_n_Eulerian_chains}, we conjecture the existence of other approaches for deriving Bell inequalities via Euler chains, which provide a higher violation.
\begin{acknowledgments}
  This work was financially supported by BMBF (Germany). We thank Michael Epping and Julius Korinth for inspiring discussions.
\end{acknowledgments}
\newpage
\appendix
\section{Proofs}
\label{apdx::proofs_for_X-chains}
\begin{customprop}{\ref{prop::merging_of_intersecting_fragments}}
[Merging of intersecting fragments]
\Needspace*{6\baselineskip}
Let $f_{1}$ and $f_{2}$ be two X-chain fragments and $f_{1}\cap f_{2}\not=\emptyset $, then $f_{1}\cup f_{2}$ is an X-chain fragment.
\begin{proof}
Let $\xi $ be an X-chain such that $\xi \cap \left( f_{1}\cup f_{2}\right)\not=\emptyset $.
Let $v\in $ $\xi \cap \left( f_{1}\cup f_{2}\right) $, then $v\in $ $\xi \cap f_{1}$ or $v\in $ $\xi \cap f_{2}$.
Without loss of generality, we assume $v\in $ $\xi \cap f_{1}$.
Since $f_{1}$ is an X-chain fragment, according to Def. \ref{def::X-chain_fragment} $\xi \cap f_{1}=f_{1}$.
That means $\xi \cap f_{1}\cap f_{2}=f_{1}\cap f_{2}\not=\emptyset $.
Therefore there exists a vertex $v^{\prime}\in f_{1}\cap f_{2}$, such that $v^{\prime }\in $ $\xi \cap f_{2}$.
Since $f_{2}$ is an X-chain fragment, $\xi \cap f_{2}=f_{2}$.
Therefore  $\xi\cap \left( f_{1}\cup f_{2}\right) =\left( \xi \cap f_{1}\right) \cup \left(\xi \cap f_{2}\right) =f_{1}\cup f_{2}$.
Hence $f_{1}\cup f_{2}$ is an X-chain fragment.
\end{proof}
\end{customprop}
\begin{customprop}{\ref{prop::criterion_for_fragment_merging_1}}
[Criterion for X-chain fragment merging I]
\Needspace*{6\baselineskip}
Let $f_{1}$ and $f_{2}$ be two disjoint X-chain fragments.
If there exists a vertex $v$, such that $N_v\subseteq f_1\cup f_2$ and the cardinalities $|N_v\cap f_1|$ and $|N_v\cap f_2|$ are both odd,
then $f_1\cup f_2$ is an X-chain fragment.
\begin{proof}
Let $\xi$ be an X-chain.
Since $N_v\subseteq f_1\cup f_2$, it holds
\begin{equation}
  N_v\cap\xi = N_v\cap(f_1\cup f_2)\cap\xi = (N_v\cap f_1\cap\xi) \cup (N_v\cap f_2\cap\xi).
\end{equation}
Due to the disjointness of $f_1$ and $f_2$, the cardinality of $ N_v\cap\xi $ is then equal to
\begin{equation}
  | N_v\cap\xi | = |N_v\cap f_1\cap\xi| + |N_v\cap f_2\cap\xi|.
\end{equation}
According to the criterion for X-chains (Theorem \ref{theorem::criterion_for_X-chains}), i.e. $|N_v\cap\xi|=0 \pmod 2$, it follows that $|N_v\cap f_1\cap\xi|$ and $|N_v\cap f_2\cap\xi|$ must have the same parity.
According to the definition of X-chain fragments (Def. \ref{def::X-chain_fragment}), it holds $f_1\cap\xi\in \{
\emptyset, f_1\}$ and $f_2\cap\xi\in \{
\emptyset, f_2\}$.
\par
Assume that $f_1\cap\xi=f_1$, then $|N_v\cap f_1\cap\xi|=|N_v\cap f_1|=1 \pmod 2$, which follows that $|N_v\cap f_2\cap\xi| = 1 \pmod 2$, and hence there exist $v\in f_2\cap \xi$, which indicates $f_2\cap\xi=f_2$.
Then in this case $(f_1\cup f_2)\cap\xi=(f_1\cup f_2)$.
\par
Assume that $f_1\cap\xi=\emptyset$, then $|N_v\cap f_1\cap\xi|= 0 \pmod 2$, which follows that $|N_v\cap f_2\cap\xi| = 0 \pmod 2$.
Therefore $f_2\cap\xi=\emptyset$, otherwise $f_2\cap\xi=f_2$ and $|N_v\cap f_2\cap\xi|=|N_v\cap f_2|=1\pmod 2$, which leads to a contradiction.
Then in this case $(f_1\cup f_2)\cap\xi=\emptyset$.
\par
As a result, $f_1\cup f_2$ is an X-chain fragment.
\end{proof}
\end{customprop}%
\begin{customprop}{\ref{prop::criterion_for_fragment_merging_2}}
[Criterion for X-chain fragment merging II]
\Needspace*{14\baselineskip}
Let $f_{1}$ and $f_{2}$ be two disjoint X-chain fragments.
If there exist two vertices $v_{1}$ and $v_{2}$, such that%
\begin{equation}
N_{v_{1}}\setminus N_{v_{2}}=f_{1} \text{ and } N_{v_{2}}\setminus N_{v_{1}}=f_{2},
\end{equation}%
and the edge number between $v_i$ and $f_i$ is odd for $i=1,2$, i.e.
\begin{equation}\label{eq::prop_criterion_for_fragment_merging_cond}
|N_{v_i}\cap f_i|=1 \pmod 2,
\end{equation}
then $f_{1}\cup f_{2}$ is an X-chain fragment.
\begin{proof}
Let $\xi $ be an X-chain containing $f_{1}$, i.e. $f_1\cap\xi=f_1$.
Since $|N_{v_1}\cap f_1|=1 \pmod 2$ and $|N_{v_1}\cap \xi|=0 \pmod 2$ (Theorem \ref{theorem::criterion_for_X-chains}),
the cardinality
\begin{equation}
| \xi\cap (N_{v_1}\setminus f_1) |=|\xi\cap N_{v_1}|-|\xi\cap N_{v_1}\cap f_1|
\end{equation}
must be odd.
Since $N_{v_{1}}\setminus f_{1} =N_{v_{1}}\cap N_{v_{2}} =N_{v_{2}} \setminus f_{2} $,
the cardinality $| \xi\cap N_{v_2}\setminus f_2 |$ is odd.
Meanwhile $|\xi \cap N_{v_2} |=| \xi\cap (N_{v_2}\setminus f_2) |+|\xi \cap f_2 |$  must be even (Theorem \ref{theorem::criterion_for_X-chains}),
hence  $|\xi \cap f_2 |=1 \pmod 2$.
This follows $f_{2} \cap \xi = f_2$ (Def. \ref{def::X-chain_fragment}).
Therefore $\left( f_{1}\cup f_{2}\right) \cap \xi =f_{1}\cup f_{2}$.
\par
Let $\xi $ be an X-chain not containing $f_1$, i.e. $f_1\cap \xi=\emptyset$.
Then according to Theorem \ref{theorem::criterion_for_X-chains}, $| \xi\cap (N_{v_2}\setminus f_2) |
=| \xi\cap (N_{v_1}\setminus f_1) | =| \xi\cap N_{v_1} |$ and $| \xi\cap N_{v_2} |$ are even.
Hence $| \xi \cap N_{v_2}\cap f_2 |=| \xi\cap N_{v_2} |-| (\xi\cap N_{v_2})\setminus f_2 |$ is also even.
It follows $\xi\cap f_2=\emptyset$, otherwise $| \xi \cap N_{v_2}\cap f_2 |=| N_{v_2}\cap f_2 |=1 \pmod 2$, which leads to contradiction.
Therefore $\left( f_{1}\cup f_{2}\right) \cap \xi=\emptyset $.
\par
As a result, $f_{1}\cup f_{2}$ is an X-chain fragment.
\end{proof}
\end{customprop}%
\begin{customcoro}{\ref{coro::X-chains_and_bias_degree_of_L_n}}
[X-chains and bias degree of linear cluster states]
\Needspace{12\baselineskip}
The X-chain generators and Z-bias degree of linear cluster states $|L_{n}\rangle$ with vertex set $V_{L_{n}}=\left\{ 1,...,n\right\}  $ (see Table \ref{table::X-chains_of_special_graphs_Ln}) depend on the parity of vertex number $n$.
\begin{enumerate}
\item If $n$ is odd, $\Gamma_{L_{n}}=\left\{  \gamma_{\mathrm{odd}}\right\}  $ with $\gamma_{\mathrm{odd}}=\left\{  1,3,...,n-2,n\right\}  $, and $\beta(|L_{n}\rangle)=2^{-\left(  n-1\right)  /2}$.

\item If $n$ is even, $\Gamma_{L_{n}}=\emptyset$, and $\beta(|L_{n}%
\rangle)=2^{-n/2}$.
\end{enumerate}
\begin{proof}
We index the vertices of linear cluster states by $\{1,...,n\}$ as shown in Table \ref{table::X-chains_of_special_graphs_Ln}.
Every vertex $i\in\left\{ 2,...,n-1\right\}  $ has degree $d_{i}=2$.
Therefore, according to Proposition \ref{prop::criterion_for_fragment_merging_1}, $\left\{
1,3\right\}  $, $\left\{  2,4\right\}  $, $\left\{  3,5\right\}  $, ...,
$\left\{  n-2,n\right\}  $ are all possible candidates for X-chain fragments.
As a result of Proposition \ref{prop::merging_of_intersecting_fragments}%
, they can be merged into $f_{\mathrm{odd}}=\left\{  1,3,5,...,\right\} $ and $f_{\mathrm{even}}=\left\{2,4,6...,\right\} $.
The vertex set $f_{\mathrm{even}}$ is not an X-chain fragment, since $|N_1\cap f_{\mathrm{even}}|=1$ (Proposition
\ref{prop::criterion_for_redundant_fragments}).
\par
If $n$ is odd, then $f_{\mathrm{odd}}=\left\{ 1,3,5,...,n\right\}  $.
For any vertex $v\in V_{L_n}$, $|N_v\cap f_{\mathrm{odd}}|=0\pmod 2$, since $|N_{2k+1}\cap f_{\mathrm{odd}}|=0$ and $|N_{2k}\cap f_{\mathrm{odd}}|=2$ for all $k=1,...,(n-1)/2$.
Therefore $\left\{  1,3,5,...,n\right\}  $ is the only X-chain generator.
According to Eq. \eqref{eq::bias_degrees_of_graph_states}, the bias degree is $\beta(|L_{n}\rangle
)=2^{-\left(  n-1\right)  /2}$.
\par
If $n$ is even, then $f_{\mathrm{odd}}=\left\{ 1,3,5,...,n-1\right\}  $.
Since $|N_n\cap f_{\mathrm{odd}}|=1$, $f_{\mathrm{odd}}$ is not an X-chain fragment (Proposition
\ref{prop::criterion_for_redundant_fragments}).
Hence there is no X-chain generator.
According to Eq. \eqref{eq::bias_degrees_of_graph_states}, the bias degree is $\beta(|L_{n}\rangle)=2^{-n/2}$.
\end{proof}
\end{customcoro}
\begin{customcoro}{\ref{coro::X-chains_and_bias_degree_of_K_n}}
[X-chains and bias degree of complete graph states]
\Needspace*{12\baselineskip}
The X-chain generators and Z-bias degree of complete graph states $|K_{n}\rangle $ with vertex set $V_{K_{n}}=\left\{ 1,...,n\right\} $ (see Table. \ref{table::X-chains_of_special_graphs_Kn}) depend on the parity of vertex number $n$.
\begin{enumerate}
\item If $n$ is odd, $\Gamma _{K_{n}}=\left\{ V_{K_{n}}\right\} $, and $\beta (|K_{n}\rangle )=2^{-\left( n-1\right) /2}$ for $n=1 \pmod 4$, while $\beta (|K_{n}\rangle )=0$ for $n=3 \pmod 4$.
\item If $n$ is even, $\Gamma _{K_{n}}=\emptyset $, and $\beta(|K_{n}\rangle )=2^{-n/2}$.
\end{enumerate}
\begin{proof}
For any vertex pair $i,j\in V_{K_{n}}$, it holds $N_{i}\setminus N_{j}=\{j\}$ and $N_{j}\setminus N_{i}=\{i\}$. According to Proposition \ref{prop::criterion_for_fragment_merging_2}, any vertex pair $\{i,j\} $ is a candidate for X-chain fragments.
As a result of Proposition \ref{prop::merging_of_intersecting_fragments}, the entire vertex set $V_{K_{n}}$ is a candidate for X-chain fragment.
\par
If $n$ is odd, every vertex $v$ has even number of neighbors $N_v=n-1$.
Therefore $|N_v\cap V_{K_{n}}|=0 \pmod 2$ for all vertices $v$, and $V_{K_{n}}$ is the only X-chain generator of the complete graph state $\ket{K_{n}}$.
If $n=1 \pmod 4$, then the edge number $|E_{K_n}|=\binom{4k+1}{2}$ with $k\in\mathbb{N}$ is even, hence the stabilizer parity $\pi_G(V_{K_n})=1$ (see Eq. \eqref{eq::G-parity_formula}),
If $n=3 \pmod 4$, then the edge number $|E_{K_n}|=\binom{4k+3}{2}$ is odd, hence one arrives at the stabilizer parity $\pi_G(V_{K_n})=-1$.
According to Eq. \eqref{eq::bias_degrees_of_graph_states}, the bias degree is $\beta (|K_{n}\rangle )=2^{-\left( n-1\right) /2}$ for $n=1 \pmod 4$, while $\beta (|K_{n}\rangle )=0$ for $n=3 \pmod 4$.
\par
If $n$ is even, every vertex $v$ has an odd number of neighbors $N_v=n-1$. That means $V_{K_{n}}$ is not an X-chain fragment.
Therefore $\Gamma_{K_{n}}=\emptyset $, and $\beta (|K_{n}\rangle )=2^{-n/2}$.
\end{proof}
\end{customcoro}
\begin{customcoro}{\ref{coro::X-chains_and_bias_degree_of_C_n}}
[X-chains and bias degree of cycle graph states]
\Needspace{12\baselineskip}
The X-chain generators and Z-bias degree of cycle graph states $|C_{n}\rangle $ with vertex set $V_{C_{n}}=\left\{1,...,n\right\} $ (see Table. \ref{table::X-chains_of_special_graphs_Cn}) depend on
the parity of vertex number $n$.
\begin{enumerate}
\item If $n$ is odd, $\Gamma _{C_{n}}=\left\{ V_{C_{n}}\right\} $, and $\beta (|C_{n}\rangle )=0$.
\item If $n$ is even, $\Gamma _{C_{n}}=\left\{ \gamma _{\mathrm{odd}},\gamma_{\mathrm{even}}\right\} $, with $\gamma_{\mathrm{odd}}=\left\{ 1,3,5,...,n-1\right\} $ and $\gamma _{\mathrm{even}}=\left\{ 2,4,6...,n\right\} $.
    The Z-bias degree is $\beta(|C_{n}\rangle )=2^{-(n-2)/2}$.
\end{enumerate}
\begin{proof}
Every vertex $i\in V_{C_{n}}$ has degree $d_{i}=2$ (see Table \ref{table::X-chains_of_special_graphs_Cn}). Similar to the proof of Corollary \ref{coro::X-chains_and_bias_degree_of_L_n}, $\gamma_{\mathrm{odd}}$ and $\gamma_{\mathrm{even}}$ are two candidates for X-chain fragments.
Besides $\gamma_{\mathrm{odd}} $ and $\gamma_{\mathrm{even}}$, $\left\{ n,2\right\} $ is also an candidate for X-chain fragment.
\par
If $n$ is odd, according to Proposition \ref{prop::merging_of_intersecting_fragments}, the three candidates can be merged into the whole vertex set $V_{C_n}=\gamma_{\mathrm{odd}} \cup \gamma_{\mathrm{even}}$.
Furthermore $N_i\cap V_{C_n}=0 \pmod 2$ for every vertex $i\in V_{C_n}$ (Theorem \ref{theorem::criterion_for_X-chains}), therefore $V_{C_{n}}$ is the only X-chain generator of $|C_{n}\rangle $.
According to Eq. \eqref{eq::bias_degrees_of_graph_states}, the stabilizer parity of $V_{C_{n}}$ is negative, therefore $\beta (|C_{n}\rangle)=0$ and $\ket{C_{n}}$ is a Z-balanced graph state.
\par
If $n$ is even, $\gamma_{\mathrm{odd}} $ and $\gamma_{\mathrm{even}}$ cannot be merged into one X-chain fragment.
Since the X-chain criterion holds for both $\gamma_{\mathrm{odd}} $ and $\gamma_{\mathrm{even}}$, $\gamma_{\mathrm{odd}} $ and $\gamma_{\mathrm{even}}$ are the X-chain generators.
According to Eq. \eqref{eq::bias_degrees_of_graph_states}, the bias degree is $\beta(|C_{n}\rangle)=2^{-(n-2)/2}$.
\end{proof}
\end{customcoro}
\begin{customthm}{\ref{theorem::MP_Bell_ineq_n_Eulerian_chains}}
[Bell inequalities for graph states and Euler chains]
Let $|G\rangle$ be a graph state.
If $\xi$ is an Euler chain of $\ket{G}$ with negative stabilizer parity, the following multipartite Bell operator%
\begin{equation}
B_{\xi}=\sum_{i\in\xi}g_{i}+s_{G}^{(\xi)}
\end{equation}
with $g_{i}=\sigma_{x}^{(i)}\sigma_{z}^{(N_{i})}$ is dichotomic.
Its expectation value in a local hidden variable model is bounded by
\begin{equation}
  -|\xi|+(-1)^{|\xi|+1} \leq \braket{B_{\xi}}_{LHV} \leq |\xi|-1,
\end{equation}
while its quantum bound $\left\vert \xi\right\vert +1$ is reached by the graph state $\ket{G}$.
\begin{proof}
  The expectation value $\braket{B_{\xi}}$ in a local hidden variable model is calculated by
  \begin{equation}
    \braket{B_{\xi}}_{\mathrm{LHV}} = \int \left( \sum_i  \tilde{g}_i(\lambda) + \tilde{s}_{G}^{(\xi)}(\lambda) \right) p_{\lambda} d\lambda,
  \end{equation}
  where $\tilde{g}_i(\lambda)$ and $\tilde{s}_{G}^{(\xi)}(\lambda)$ are the values of the operators $g_i$ and $s_{G}^{(\xi)}$ assigned by the local hidden variable $\lambda$, respectively, and $p_{\lambda}$ is the probability of $\lambda$.
  Since $\xi$ is an Euler chain with negative parity, i.e. $\xi\cap c_{\xi}=\emptyset$ and $\pi_G(\xi)=-1$,  according to Eq. \eqref{eq::induced_stabilizer_math_formula}, it holds
  \begin{equation}
    \tilde{s}_{G}^{(\xi)}(\lambda) = -\tilde{\sigma}_{X}^{(\xi)}(\lambda)\tilde{\sigma}_{Z}^{(c_{\xi})}(\lambda),
  \end{equation}
  with $\tilde{\sigma}_{X}^{(\xi)}(\lambda)$ and $\tilde{\sigma}_{Z}^{(c_{\xi})}(\lambda)$ being the operator values assigned by $\lambda$.
  Due to the commutative relationship between the scalar values $\tilde{\sigma}_X(\lambda)$ and $\tilde{\sigma}_Z(\lambda)$,
  the product of $\tilde{g}_{i}(\lambda)$ is calculated to
  \begin{equation}
    \prod_{i\in \xi} \tilde{g}_{i}(\lambda) = \tilde{\sigma}_{X}^{(\xi)}(\lambda)\tilde{\sigma}_{Z}^{(c_{\xi})}(\lambda).
  \end{equation}
  Hence, the expectation value $\braket{B_{\xi}}_{\mathrm{LHV}}$ is bounded by the extremal values of
  \begin{equation}
    \sum_{i\in \xi}\tilde{g}_{i}(\lambda)-\prod_{i\in \xi} \tilde{g}_{i}(\lambda),
  \end{equation}
  i.e. by the maximum and minimum of
  \begin{equation}
    a_1+\cdots+a_{|\xi|}-a_1a_2\cdots a_{|\xi|}
  \label{eq::proof_theorem_MP_Euler_chain_1}
  \end{equation}
  with $a_i=\pm1$.
  Since $1-a_1\cdots a_{k-1}\geq0$, by setting $a_k=+1$ one obtains the upper bound of the expression
  \begin{equation}
    a_1+\cdots+a_{k-1}+a_{k}-a_1\cdots +a_{k-1}a_{k} \leq a_1+\cdots+a_{k-1}-a_1\cdots a_{k-1}+1.
  \end{equation}
  On the other hand, since $1\pm a_1\cdots a_{k-1}\geq0$, by setting $a_k=-1$ one obtains the lower bound
  \begin{equation}
    a_1+\cdots+a_{k-1}+a_{k}\pm a_1\cdots +a_{k-1}a_{k} \geq a_1+\cdots+a_{k-1}\mp a_1\cdots a_{k-1}-1.
  \end{equation}
  Iteratively bounding the Eq. \eqref{eq::proof_theorem_MP_Euler_chain_1} by $a_k=\pm1$ with $k=|\xi|,...,1$, one obtains
  \begin{equation}
    -|\xi|+(-1)^{|\xi|+1} \leq a_1+\cdots+a_{|\xi|}-a_1a_2\cdots a_{|\xi|} \leq |\xi|-1.
  \end{equation}
  On the other hand, since $B_{\xi}$ is a sum of stabilizer operators of the graph state $\ket{G}$, the quantum mechanical upper bound of $\braket{B_{\xi}}$ is given by the number of stabilizers $|\xi|+1$. This bound can be reached by the state $\ket{G}$.
\end{proof}
\end{customthm}
%
    %
\bibliographystyle{iopart-num}
\bibliography{The_search_of_X-chains}
    %

\end{document}